# Efficiency at maximum power of thermoelectric heat engines with the symmetric semiconductor superlattice

Xiaoguang Luo, Hexin Zhang, Dan Liu, Nannan Han, Dong Mei, Jinpeng Xu, Yingchun Cheng[*], and Wei Huang[*]

*Frontiers Science Center for Flexible Electronics (FSCFE), Shaanxi Institute of Flexible Electronics (SIFE) & Shaanxi Institute of Biomedical Materials and Engineering (SIBME), Northwestern Polytechnical University (NPU), 127 West Youyi Road, Xi'an 710072, China.*

**Abstract**

Efficiency at maximum power (EMP) is a very important specification for a heat engine to evaluate the capacity of outputting adequate power with high efficiency. It has been proved theoretically that the limit EMP of thermoelectric heat engine can be achieved with the hypothetical boxcar-shaped electron transmission, which is realized here by the resonant tunneling in the one-dimensional symmetric InP/InSe superlattice. It is found with the transfer matrix method that a symmetric mode is robust that regardless of the periodicity, and the obtained boxcar-like electron transmission stems from the strong coupling between symmetric mode and Fabry-Pérot modes inside the allowed band. High uniformity of the boxcar-like transmission and the sharp drop of the transmission edge are both beneficial to the maximum power and the EMP, which are optimized by the bias voltage and the thicknesses of barrier and well. The maximum power and EMP are extracted with the help of machine learning technique, and more than 95% of their theoretical limits can both be achieved for smaller temperature difference, smaller barrier width and larger well width. We hope the obtain results could provide some basic guidance for the future designs of high EMP thermoelectric heat engines.

**1. Introduction**

On the way to ease the world energy shortage, thermoelectric technology has been considered as a promising strategy to extract the useful electric power from the sustainable or waste heat sources, such as the solar, geothermal energy, industrial waste heat, etc [1-4]. Towards a thermoelectric heat engine (TEHE), electrons driven by the temperature gradient can be drifted against the electrochemical potential gradient, and generate power. Thanks to the configuration without moving part, TEHEs are welcomed in many application scenarios due to the working stability and reliability [5]. Particularly, the figure of merit of a thermoelectric material can be boosted effectively with the decrease of its dimension due to the adjustable electrical conductivity and the great reduction of

---

thermal conductivity [6-9], which becomes a golden key for the exploitation of miniaturized or integratable TEHEs.

Efficiency $\eta$ and power $\mathcal{P}$ are two key performance parameters of a TEHE. At cryogenic temperature, the phonon effects will be strongly weakened, and the adjustable electrical conductivity plays a significate role on the TEHE performance. An effective way to improve the efficiency is adopting the energy-dependent electron transmission [10], which might be realized by the quantum confinement in quantum dot, quantum well, nanowire, superlattice etc. The maximum efficiency of a TEHE has been theoretically obtained with the delta-shaped transmission [10-12], i.e. $\mathcal{T}(\varepsilon) = \delta(\varepsilon - \varepsilon_0)$, where $\varepsilon$ and $\varepsilon_0$ are the electron energy and allowed energy level, respectively. The maximum efficiency can even reach the universal upper bound (the Carnot efficiency $\eta_C = 1 - T_C/T_H$) when the TEHE works reversibly, with the temperature of cold/hot reservoir $T_{C/H}$. However, the power is very small and even vanishes with the increase of efficiency because of the very few working electrons. To search the optimum working conditions, a widely accepted trade-off proposal is to find the efficiency at maximum power (EMP). A few decades ago, an EMP upper bound called Curzon-Ahlborn efficiency $\eta_{CA} = 1 - \sqrt{1 - \eta_C}$ had been found from the endoreversible Carnot heat engine [13]. It is conformed that the Curzon-Ahlborn efficiency is not the universal limit (e.g. surpassed by the EMP upper bound of TEHE [14, 15]), which is ensured in the range of $\eta_C/2$ to $\eta_C/(2 - \eta_C)$ [16]. Similar to the maximum efficiency, the EMP upper bound of TEHE has also been obtained under the condition of delta-shaped transmission, where the heat and electron fluxes couple strongly with each other [15]. The power therein, however, is still very small due to the ultra-narrow transmission window. For example, a single quantum dot TEHE operating close to the EMP upper bound had recently been achieved in experiments [17, 18], where the nanoscaled quantum dot was fabricated by the InP/InSe/InP heterostructure embedded in an InSe nanowire. Great energy selection from quantum confinement effect for tunneling electrons in this quantum dot enhances the performance of TEHE. The high EMP (close to Curzon-Ahlborn efficiency) is mainly ascribed to cotunneling effect in such a large quantum dot [19], with the sacrifice of small power. Thus, those uncoupled scenarios with wider transmission window may be of more practical significance in some way. In our previous work [20], the EMP with finite power has been found when the boxcar function (or uniform function) is bluntly employed as the electron transmission function. Subsequently, this EMP had been demonstrated by Whitney to be the upper bound at the finite power through a quantum TEHE model [21, 22], where the author had also provided a hypothetical implementation of quantum dot chain to realize the boxcar-like transmission.

So far, a boxcar-like transmission has been observed experimentally in the GaAs/AlGaAs superlattice that sandwiched by two antireflection coating layers (AlGaAs layers) [23]. A number of recent works have also demonstrated that the thermoelectric performance can definitely be improved with the antireflection coating layers [24-26]. However, the formation of the boxcar-like transmission is still not clarified, and the evaluation protocols on the EMP is not yet proposed. In this paper, one-dimensional InP/InSe modified Kronig-Penney superlattice (a typical textbook example), is used for demonstrating the formation. Through the transfer matrix method [27-31], symmetry mode and Fabry-Pérot modes are distinguished for the electron transmission. The strong coupling of these two kinds of modes then results in the boxcar-like transmission. The EMP of the TEHE is then evaluated by the standard of the theoretical upper bound, after defining the full width at half maximum of the boxcar-like transmission spectrum as the energy width of the transmission window. It is found both maximum power and EMP are significantly dependent on the uniformity and the slope of the boxcar-like function, and both of them can reach beyond 95% of theoretical limits at the specific structure parameters.

**2. Boxcar-like shaped transmission**

Figures 1(a-c) show the unit cell of one-dimensional InP/InSe superlattice with the lattice constant $a = b_1 + w + b_2$. The InSe layer with width $w$ acts as the well in each unit cell, and the InP layers with total width $b = b_1 + b_2$ display as the barriers with a filling ratio of $\alpha = b/a$. The parameters of barrier height and electron effective mass are adopted from ref [32], i.e., $\phi = 0.57$ eV, $m^*_{InAs} = 0.023m_0$, $m^*_{InP} = 0.08m_0$, where $m_0$ is the effective mass of a free electron in vacuum. We define the symmetric factor as $\beta = b_2/b$, and $0 \leq \beta \leq 0.5$ based on the mirror symmetry. This superlattice will become the common Kronig-Penney superlattice if $\beta = 0$. Focus on the electrons, the transverse energy level and conduction band edge can be offset by adjusting the electrochemical potential, therefore the energy of an electron is expressed as $\varepsilon = \hbar^2 k_x^2/2m^*$, where $\hbar$ is the reduced Plank constant, $k_x$ is the wavevector along the superlattice and $m^*$ is the effective mass of an electron in the semiconductor.

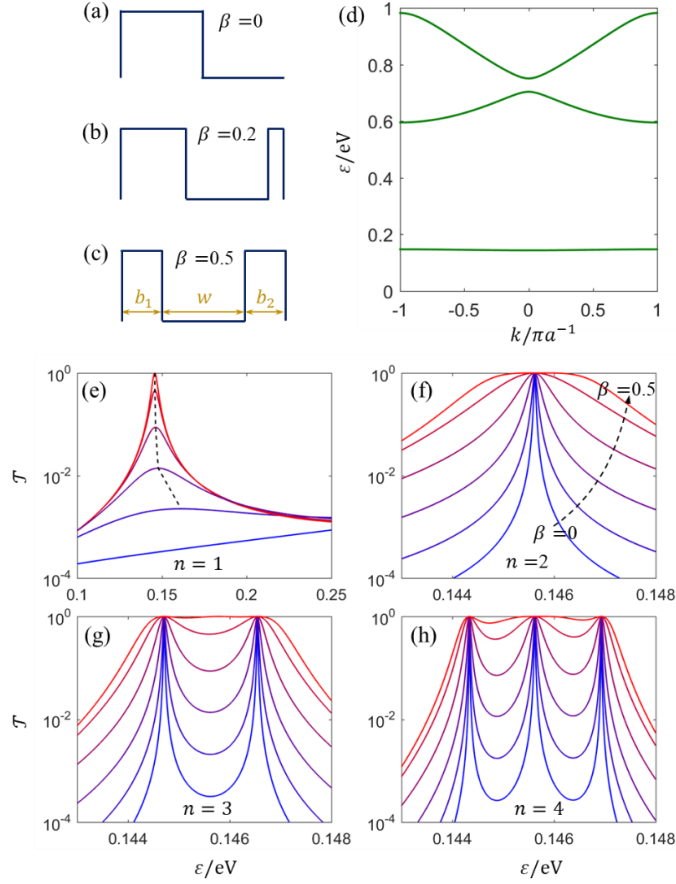

**Figure 1** (a-c) The schematic diagrams of potential profile of a unit cell of our symmetric superlattice for $\beta = 0$, $\beta = 0.2$, and $\beta = 0.5$, respectively. (d) The band structure of the symmetric superlattice for the electron energy $\varepsilon \leq 1$ eV. (e-h) The electron transmission spectrum at different $\beta$ for the periodicity $n = 1$, $n = 2$, $n = 3$, and $n = 4$, respectively. The lattice constant $a = 10$ nm and the filling ratio $\alpha = 0.5$

The electron transmission behavior in the superlattice can be described by using the transfer matrix technique [27-30]. During the calculations, the barrier structure should be divided into $N$ slices of discrete rectangle potential region, with the potential of each slice being the average value of its two sides. From the principle of finite-element approach, more precise results can be achieved with larger $N$. At any position $x$ inside the superlattice, the wavefunction can be calculated from the stationary Schrödinger equation in the framework of effective mass approximation [31],

$$\left[-\frac{\hbar^2}{2}\frac{d}{dx}\frac{1}{m^*(x)}\frac{d}{dx} + \phi(x)\right]\psi(x) = \varepsilon\psi(x) \quad (1)$$

where $m^*(x)$ and $\phi(x)$ are the electron effective mass and the potential at position $x$, respectively. Then, the general expression of wavefunction in the $j$th slice can be expressed as

$$\psi_j(x) = a_j e^{ik_j x} + b_j e^{-ik_j x} \quad (2)$$

where $k_j = \sqrt{2m_j^*(\varepsilon - \phi_j)}/\hbar$ is the wavevector, $m_j^*$ and $\phi_j$ are the electron effective mass and the average potential in $j$th slice, respectively. The two parts of the wavefunction in Eq. (2) strand for two traveling waves in the opposite direction, which can be regarded as two elements in the derivation of the $2 \times 2$ transfer matrix. The segmentation treatment on the potential implies that two different kinds of transfer matrices should be addressed, used to describe the propagation inside the same slice and the transmission through the interface between two neighboring slices. With the boundary conditions from the continuity of probability density and probability current, one can obtain two transfer matrices as

$$P_j = \begin{pmatrix} e^{-ik_j d_j} & 0 \\ 0 & e^{ik_j d_j} \end{pmatrix} \tag{3a}$$

$$T_{j,j+1} = \frac{1}{2}\begin{pmatrix} 1 + k_{j+1}m_j/k_j m_{j+1} & 1 - k_{j+1}m_j/k_j m_{j+1} \\ 1 - k_{j+1}m_j/k_j m_{j+1} & 1 + k_{j+1}m_j/k_j m_{j+1} \end{pmatrix} \tag{3b}$$

and the final $2 \times 2$ transfer matrix can be calculated as

$$M = \cdots T_{j-1,j} \cdot P_j \cdot T_{j,j+1} \cdot P_{j+1} \cdots \tag{4}$$

According to the definition of probability current, the transmission probability is

$$\mathcal{T}_{1,N} = \frac{m_1 k_N}{m_N k_1}\left|\frac{1}{M_{11}}\right|^2 \tag{5}$$

For any given bias voltage $V$ applied on the barrier structure, the transmission probability $\mathcal{T}(\varepsilon, V)$ of the electrons with energy $\varepsilon$ can be obtained. Notice that the more rigorous approach to model the potential profile is the self-consistent method coupled with the Poisson's equation after considering the space charge effect [33, 34]. However, the deviations from our model are expected to be insignificant because we focus on the case under small bias voltage $V$, with assumption of linearly tilted potential profile. In fact, this method can be adopted to handle with arbitrary potential profile.

From the Bloch theorem, one can calculate the band structure of the superlattice through

$$M_u \vec{\psi} = e^{ika}\vec{\psi} \tag{6}$$

if the periodicity $n \to \infty$, where $M_u$ is the transfer matrix of a unit cell, $\vec{\psi}$ is a vector represents the two traveling waves, $k$ here is the Bloch wavevector. Figure 1(d) shows the band structure when the lattice constant $a = 10$ nm and the filling ratio $\alpha = 0.5$. It is found the band structure is changeless no matter what $\beta$ is chosen, probably because the two outmost layers do not affect the electron states of the lattice at infinite periodicity. In this paper, we focus on the first miniband that bellows the barrier height, where the electrons are transmitted through tunneling.

The transmission probability for different periodicity $n$ and filling ratio $\beta$ at $a = 10$ nm is calculated through the transfer matrix method, with the results shown in Figures 1(e-h). For the case of $n = 1$, the resonance appears when $\beta \neq 0$, and the peak value increases for larger $\beta$. Notice that not all the peaks locate in the allowed band. The unitary peak corresponds to the coherent resonant tunneling when $\beta = 0.5$, which is clearly due to the symmetry of the potential profile. The peak energy (~ 0.1458 eV) is robust even at different $n$. Thus, this mode is reasonably called as *symmetry mode*. For the cases of $n > 1$, $n - 1$ more unitary peaks can be observed in the transmission spectrum. Different from the symmetry mode, this kind of modes are robust at different $\beta$. The spatial phase evolution is adopted to unveil the nature of these modes, as shown the case of $n = 4$ in Figure 2. Other than the symmetry mode (i.e., $p_3$), the phase change of three other peaks (i.e., $p_1$, $p_2$, and $p_4$) are $\pi$, $2\pi$, and $3\pi$, respectively, which are in accordance with the Fabry-Pérot resonance condition

$$kna = z\pi \tag{7}$$

where $z = 1, 2, \cdots, n-1$ at the first miniband. Thus, this kind of modes can be called as *Fabry-Pérot modes*. Combined with Figures 1(f-h), boxcar-like transmission is formed when $\beta = 0.5$, which is evidently attributed to strong coupling of symmetry mode and Fabry-Pérot modes in the symmetric superlattice.

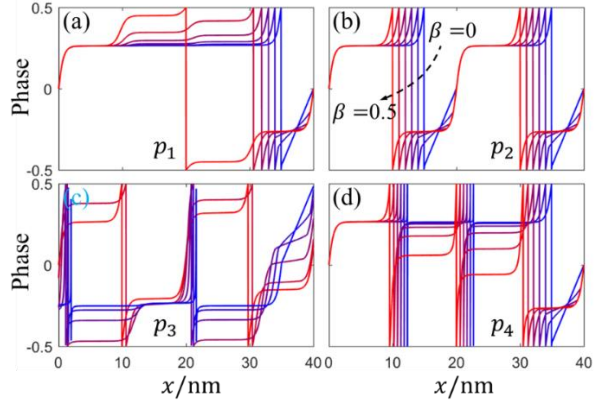

**Figure 2** (a-d) The spatial phase evolution of Figure 1(h) for the four peaks $p_1$, $p_2$, $p_3$, and $p_4$, the corresponding energy are 0.1443 eV, 0.1456 eV, 0.1458 eV, and 0.1469 eV, respectively.

## 3. The TEHE model

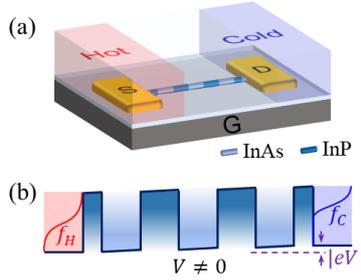

**Figure 3** (a) The schematic diagram of the TEHE model with the symmetric InP/InAs superlattice when $n = 3$ and $\beta = 0.5$, where S, D and G denote the source, drain and gate electrodes respectively. (b) The potential profile of the superlattice when the bias voltage $V \neq 0$, $f_{H/C}$ denotes the Fermi distribution of hot/cold reservoir.

Figure 3(a) shows the TEHE model with the nanowire of InP/InSe symmetric superlattice, which can be prepared by epitaxial growth [17, 35]. The periodicity is specified as $n = 3$ in this study, and similar results can certainly be obtained for the others. The contact of InAs with source/drain electrode is concepted as hot/cold reservoir, with the temperature $T_{H/C}$ that can be tuned and determined with the push-pull configuration in experiments [17]. The electrons in reservoirs follows the Fermi-Dirac distribution, and the electrochemical potentials $\mu_{H/C}$ of hot/cold reservoir can be directly tailored by the electrostatic gating. For a small bias voltage $V$, the potential profile of the superlattice will be tilted along the nanowire, as shown in Figure 3(b).

The current of the device is counted by the Landau formula [34] that expressed as

$$\mathcal{I} = \frac{2q}{h}\int_0^\infty (f_H - f_C)\mathcal{T}(\varepsilon,V)d\varepsilon \tag{8}$$

where $f_v = \{\exp[(\varepsilon - \mu_v)/k_B T_v] + 1\}^{-1}$ is the Fermi-Dirac function, $v = H$ or $C$, $q$ is the elementary charge, $h$ is Plank constant, $k_B$ is the Boltzmann constant, and "2" counts the spin of electrons. The reservoir will release or absorb an average amount of heat $\varepsilon - \mu_{H/C}$ once an electron is injected in or drawn out. Therefore, the heat flux can be calculated by

$$\dot{Q}_v = \frac{2}{h}\int_0^\infty (\varepsilon - \mu_v)(f_H - f_C)\mathcal{T}(\varepsilon,V)d\varepsilon \tag{9}$$

without considering the heat leakage. Then, the power and the efficiency of the heat engine are

$$\mathcal{P} = \dot{Q}_H - \dot{Q}_C \tag{10}$$

and

$$\eta = \frac{\mathcal{P}}{\dot{Q}_H} \tag{11}$$

respectively.

**3. The performance of TEHE**

The transmission spectrum is an important factor that related to the TEHE performance. It has been reported that the performance of thermoelectric devices can be greatly enhanced by the boxcar-like transmission [24, 26]. And the EMP upper bound of a TEHE can be obtained with the ideal boxcar transmission function [15, 20, 22]

$$\mathcal{T}(\varepsilon,V) = \begin{cases} 1 & \varepsilon_l \leq \varepsilon \leq \varepsilon_u \\ 0 & \text{elsewhere} \end{cases} \tag{12}$$

where $\varepsilon_{l/u}$ is the lower/upper energy edge of the transmission window. Notice that this rigorous boxcar transmission is hard to be realized in experiments or even in theory. To evaluate the electron transmission property in our model, the full width at half maximum of the transmission spectrum around the first miniband is regarded as the energy width of the transmission window, written by $\Delta\varepsilon = \varepsilon_u - \varepsilon_l$. Figure 4(a) shows the boxcar-like transmission formed at the first miniband under no bias. The location and the energy width of the transmission window is mainly determined by the barrier and well widths, respectively. Red-shift will be brought with the broadening well because of the increasing electron resonant wavelength, and narrower transmission window will be induced with broader barrier from the tunneling property. More details related to the transmission window are shown in Figure A1, which depicts clearly the energy width is narrowed with the increase of both barrier and well widths.

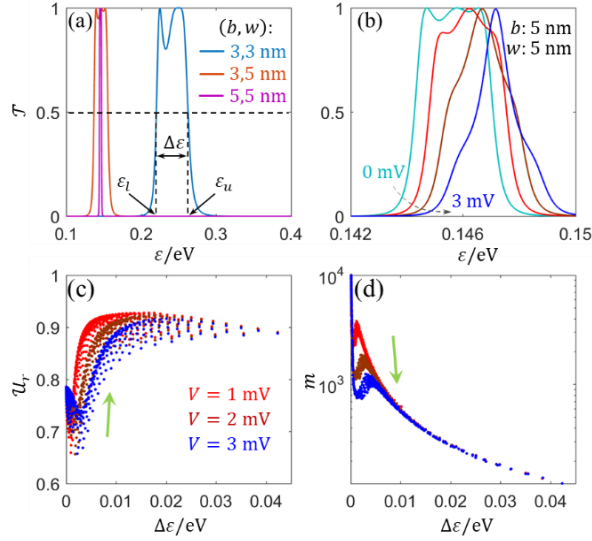

**Figure 4** (a) The electron transmission around the first miniband of the symmetric InP/InAs superlattice without bias voltage, where the energy width of the transmission window is defined as the full width at half maximum. (b) The changes of the transmission spectrum with respect to the bias voltage $V$. (c) The uniformity $\mathcal{U}_r$ and (d) the slope $m$ of the symmetric superlattice with respect to the energy width of transmission window, the data is extracted from the calculated results when the barrier width and the well width are both from 3 nm to 8 nm with the step of 0.2 nm. The green arrows point roughly out the coupling regions.

Under the bias voltage, the potential profile tilts. The asymmetry in potential destroys the coherent resonance condition, and finally results in the decoupling phenomenon. Figure 4(b) shows two shoulder peaks (Fabry-Pérot modes) are more sensitive to the bias and are reduced more quickly than the middle main peak (symmetric mode). The disappearing of Fabry-Pérot modes and the narrowing of transmission peak of symmetric mode should be responsible for the decoupling. With the increase of the bias voltage, the potential tilts up more significantly, causing the blue-shift of transmission window and the decrease of peak values as well. As the decoupling goes further, the width of the transmission window narrows abruptly, as shown the case of $V = 3$ mV in Figure 4(b), and the boxcar-like shape is remarkably destroyed, which is not interested here. The coefficient of uniformity and slope are then introduced to describe the property of the boxcar-like transmission, which are defined as

$$\mathcal{U}_r = \int_{\varepsilon_l}^{\varepsilon_u} \frac{\mathcal{T}(\varepsilon,V)}{\Delta\varepsilon} d\varepsilon \quad \text{and} \quad m = \frac{1}{\varepsilon_{0.1}^* - \varepsilon_u} \tag{13}$$

respectively, where $\varepsilon_{0.1}^*$ is located around the upper energy edge of the transmission window when $\mathcal{T}(\varepsilon_{0.1}^*, V)$ is 10% of the maximum value. Notice that $0 < \mathcal{U}_r \leq 1$ and $m > 0$. For the ideal boxcar transmission, $\mathcal{U}_r = 1$ and $m \to \infty$.

Figures A2 and A3 show the numerical results of the uniformity and the slope when $b$ and $w$ vary from 3 nm to 8 nm with the step of 0.2 nm. Two different regions are formed under the bias voltage, observably from the uniformity. The two dividing regions are relevant to the degree of decoupling, and thus named coupling region and decoupling region for simplicity. Taking $b = w = 5$ nm as an example, the cases at $V = 0.1$ eV and $V = 0.3$ eV belong to the coupling region and the decoupling region, respectively. Two different regions can also be clarified by the transmission spectrum that depicted in Figure 4(b). Figures 4(c) and (d) show the uniformity and slope with respect to $\Delta\varepsilon$, where the decoupling regions can also be distinguished at small $\Delta\varepsilon$. In the coupling regions, $\mathcal{U}_r$ increases initially and then decreases gently with the increase of $\Delta\varepsilon$, while $m$ seems always decreases. As a result, an optimum region of $\Delta\varepsilon$ exists there for better boxcar lineshape of the electron transmission.

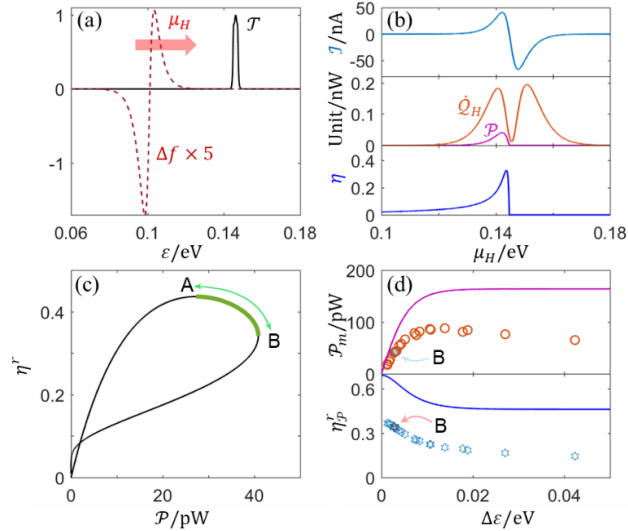

**Figure 5** (a) The strategy to extract the performance parameters by increasing the electrochemical potential of hot reservoir $\mu_H$, where $\Delta f = f_H - f_C$, $b = w = 5$ nm, $T_H = 40$ K, $T_C = 10$ K and $V = 1$ mV. (b) The current $\mathcal{J}$, power $\mathcal{P}$, heat flux from hot reservoir $\dot{Q}_H$, and efficiency $\eta$ as functions of $\mu_H$. (c) The relative efficiency $\eta^r = \eta/\eta_C$ versus power from the data of (b), A and B mark the maximum efficiency and maximum power, respectively, and the green line indicates an optimum working region. (d) The maximum power $\mathcal{P}_m$ and the correspondingly relative efficiency at maximum power $\eta_\mathcal{P}^r$ with respect to the width of the transmission window, when $b$ and $w$ increase from 3 nm to 8 nm with the step of 1 nm. The solid lines are from the ideal boxcar transmission, the symbols are from the boxcar-like transmission that working in the coupling region.

According to Eqs. (8-11), performance parameters of the TEHE can be calculated with appropriate numerical method, including the current $\mathcal{J}$, the heat flux $\dot{Q}_{H/C}$, the power $\mathcal{P}$, and the efficiency $\eta$. Two facts should be noted before calculations. One is that the lineshape of $\Delta f = f_H - f_C$ with respect to the electron energy $\varepsilon$ is unchanged at the given bias voltage and temperatures, regardless of the electrochemical potential $\mu_H$. The other is that the transmission spectrum $\mathcal{T}(\varepsilon, V)$ is absolutely determined at the specific bias voltage and potential profile. Subsequently, the strategy of the calculation based on the variable $\mu_H$ is proposed in Figure 5(a). All performance parameters can be investigated with respect to $\mu_H$. With the increase of $\mu_H$ around the first miniband, the net electron flux will initially flow from hot to cold reservoir, and then in the opposite direction. The coupled heat flux can be manipulated by the net electron flux. Figure 5(b) shows the numerical results at $V = 1$ mV when $b = w = 5$ nm. It is found both the power and the efficiency can reach the maximum values, $\mathcal{P}_m$ and $\eta_\mathcal{P}^r$, respectively, at different $\mu_H$. With the simple trade-off optimization [36, 37], an optimum region of $0.1423 \text{ eV} \leq \mu_H \leq 0.1437 \text{ eV}$ is obtained, as depicted by the bold green line in Figure 5(c).

Towards the EMP, ~34.12% of the Carnot efficiency is achieved in this TEHE, as labeled "B" in Figure 5(c). Unquestionably, $\eta_\mathcal{P}^r$ is very dependent on the working conditions, such as the temperatures, the bias voltage, and the potential profile of the superlattice. One then may ask whether there is an EMP upper bound, and how to achieve this high efficiency? Actually, these questions have already been settled in Ref. [20, 22]. The key factor is the ideal boxcar transmission function, i.e., Eq. (12). Our previous work [20] has been reported that the maximum power $\mathcal{P}_m^+$ and the corresponding EMP upper bound $\eta_\mathcal{P}^{r+}$ of a TEHE can be found with the ideal boxcar transmission at the given $T_H$, $T_C$, and $\Delta\varepsilon$, more details demonstrated in Appendix B. The ratios of $\mathcal{R}_\mathcal{P} = \mathcal{P}_m/\mathcal{P}_m^+$ and $\mathcal{R}_\eta = \eta_\mathcal{P}^r/\eta_\mathcal{P}^{r+}$ are chosen to evaluate the property of our TEHE. Figure 5(d) shows the maximum power and EMP in the coupling region, where $\Delta\varepsilon$ is extracted from the calculated results when $b$ and $w$ increase from 3 nm to 8 nm with the step of 1 nm. Both maximum power and EMP are much smaller than the ideal upper bounds (both $\mathcal{R}_\mathcal{P}$ and $\mathcal{R}_\eta$ are less than 60%, as depicted in Figure B1), although $\mathcal{U}_r$~0.9 and $m$~1500 around $\Delta\varepsilon = 5$ meV.

## 4. Optimization for EMP

To find out the reason, one need to investigate the optimization strategy for the upper bounds. The power is maximized by the condition of $\frac{\partial \mathcal{P}}{\partial \gamma_{H/C}} = 0$ at the given temperature $T_{H/C}$ and energy width of the transmission window $\Delta\varepsilon$, where the dimensionless scaled energies are defined as

$$\gamma_{H/C} = \frac{\varepsilon_l - \mu_{H/C}}{k_B T_{H/C}} \qquad (14)$$

In other words, the bias voltage $V = (\mu_C - \mu_H)/e$ is also involved into the power maximization. During the optimization, $\gamma_{H/C}$ can be obtained with respect to $\Delta\varepsilon$ at the given temperatures. One can also figure out the optimum energy difference $\varepsilon_l - \mu_{H/C}$ according to the Eq. (14), and finally identify the optimum bias voltage $V_{opt}$, as seen the case in Figure 6(a) when $T_H = 40$ K and $T_C = 10$ K. It implies the optimum bias voltage is ~3 mV, rather than the 1 mV that used in Figure 5. Therefore, one key point to approach the EMP upper bound is to choose the optimum bias voltage $V_{opt}$. Figure 6(b) displays the relationship between $V_{opt}$ and $\Delta\varepsilon$ at different $T_C$ when $T_H = 40$ K. $V_{opt}$ decreases with $\Delta\varepsilon$ and tends to a constant value when $\Delta\varepsilon > 0.01$ eV, e.g., $V_{opt}$~3 mV, 2 mV, 1 mV for $T_C = 10$ K, 20 K, 30 K.

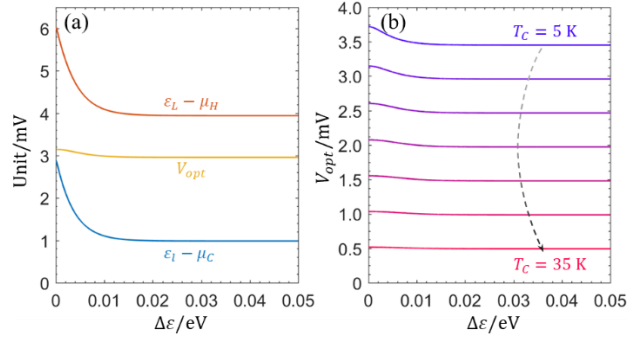

**Figure 6** (a) For the ideal boxcar transmission, the energy difference $\varepsilon_l - \mu_{H/C}$ and the optimum bias voltage $V_{opt}$ with respect to $\Delta\varepsilon$ at maximum power, here $T_H = 40$ K and $T_C = 10$ K. (b) The optimum bias voltage is very dependent on the temperature $T_C$ when $T_H = 40$ K.

We next verify the significance of the optimum bias voltage in the power maximization. The bias voltages are chosen as $V = 3$ mV, $2$ mV, $1$ mV for $T_C = 10$ K, $20$ K, $30$ K, respectively. Figure 7 shows the maximum power and EMP in the coupling region, where $\Delta\varepsilon$ is extracted when $b$ and $w$ increase from 3 nm to 8 nm with the step of 1 nm. The more suitable bias voltages significantly improve the maximum power and EMP of our TEHE. About 90% of both $\mathcal{R}_\mathcal{P}$ and $\mathcal{R}_\eta$ is obtained around $\Delta\varepsilon = 5$ meV, as seen in Figure B2. Especially, more than 95% is achieved for the case at $V = 1$ mV and $T_C = 30$ K, which are very close to the upper bounds. Notice that they cannot reach the upper bounds owing to the nature of resonant transmission. Higher ratios are realized with the smaller temperature difference and the smaller bias voltage, suggesting that the upper bounds are more easily achieved under the linear thermoelectric response. With the increase of $\Delta\varepsilon$, it is found $\mathcal{R}_\mathcal{P}$ increases initially and then decreases, implying the uniformity $\mathcal{U}_r$ and the slope $m$ affect the maximum power significantly in small and large energy width of the transmission window, respectively. In fact, Figure 6(a) shows that the optimum bias voltages are not always constant with $\Delta\varepsilon$. Thus, the optimum maximum power and EMP can be further optimized by using the machine learning technique (more details described in Appendix B), as seen by the cyan dots in Figure 7.

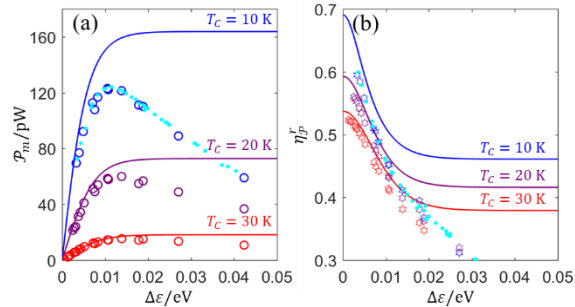

**Figure 7** (a) The maximum powers and (b) the corresponding EMP for the ideal boxcar transmission (solid line) and our boxcar-like transmission (symbols), the cyan dots denote the machine leaning results. The temperature of hot reservoir is fixed as $T_H = 40$ K, and the bias voltage $V = 3$ mV, $2$ mV, and $1$ mV for $T_C = 10$ K, $20$ K, and $30$ K, respectively.

## 5. Conclusions

The one-dimensional symmetric InP/InSe superlattice is adopted to realize the boxcar-like electron transmission, because of the coupling of symmetry mode and Fabry-Pérot modes. Boosted significantly by high uniformity of the transmission window and sharp drop of the window edge, the maximum power and the EMP of a TEHE can be improved dramatically, and even close to the theoretical limits under the optimum bias voltage and the optimum widths of well and barrier that can be obtained with the machine learning technique. More than 95% of the upper bounds for both maximum power and EMP are obtained for the temperatures of $T_H = 40$ K, $T_C = 30$ K. The results also show that it is more likely to approach the upper bounds for smaller temperature difference, smaller barrier width and larger well width, which provide some specific guidance for experiments. Notice that similar results could also be obtained for other temperatures with our approach, and better accuracy will be achieved for lower temperature because of the weaker phonon effects.

**Acknowledgements**

This work was supported by the National Natural Science Foundation of China (NSFC) (Grant No. 61905198), the Natural Science Basic Research Program of Shaanxi (Program No. 2019JQ-059), the National Postdoctoral Program for Innovative Talents (No. BX20190283), the Joint Research Funds of Department of Science & Technology of Shaanxi Province and Northwestern Polytechnical University (Nos. 2020GXLH-Z-020, 2020GXLH-Z-027 and 2020GXLH-Z-029), and the Fundamental Research Funds for the Central Universities. X. Luo would like to thank Lei Ying for the fruitful discussion.

**Appendix**

**A. The electron transmission of our model**

The electron transmission spectrum in the InP/InSe symmetric superlattice at $n = 3$ is calculated through the transfer matrix method, with the width of each slice $d = 0.01$ nm. Both barrier width and the well width increase from 3 nm to 8 nm during the calculation with the step of 0.2 nm. The energy width $\Delta\varepsilon$, the uniformity $\mathcal{U}_r$, and the slope $m$ are extracted from the data, as shown in Figure A1, A2, and A3, respectively.

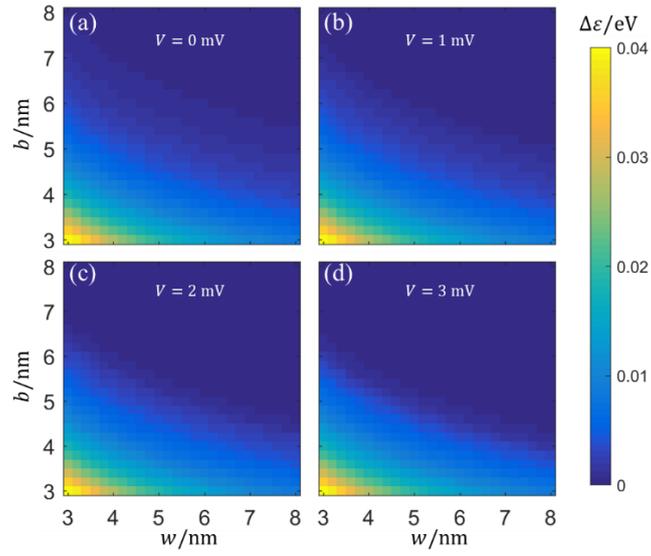

**Figure A1** The energy width of transmission window at the first miniband for the InP/InSe symmetric superlattice. The bias (a) $V = 0$ V, (b) $V = 1$ mV, (c) $V = 2$ mV, (d) $V = 3$ mV. The periodicity $n = 3$, and $b$ and $w$ are the widths of barrier and well, respectively

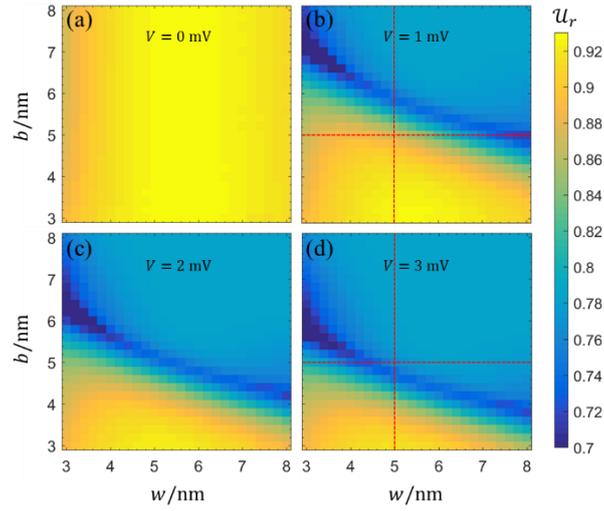

**Figure A2** The uniformity $\mathcal{U}_r$ at the bias of (a) $V = 0$ V, (b) $V = 1$ mV, (c) $V = 2$ mV, (d) $V = 3$ mV. The dotted guide lines in (b) and (d) point out the position of $b = w = 5$ nm.

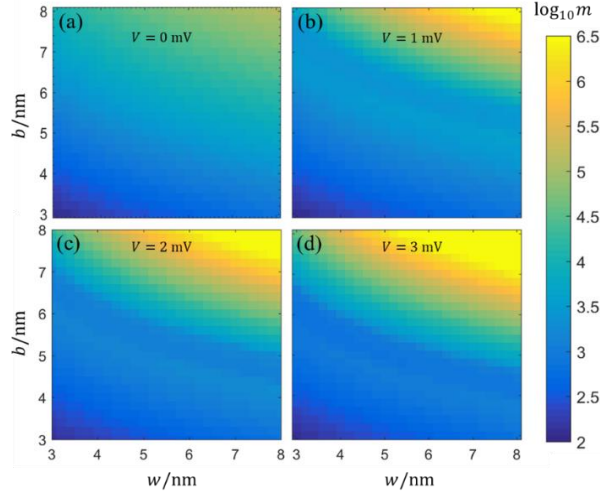

**Figure A3** The slope $m$ at the bias of (a) $V = 0$ V, (b) $V = 1$ mV, (c) $V = 2$ mV, (d) $V = 3$ mV.

## B. The efficiency at maximum power of TEHE

To obtain the EMP upper bound, the ideal boxcar transmission function of Eq. (12) should be employed during the calculation. Combined with Eqs. (9-11), the heat flux and the power can be rewritten as [15, 20]

$$\dot{Q}_H = \frac{2}{h} k_B^2 T_H^2 \{[g(e^{-\Gamma_H - \gamma_H}) - g(e^{-\gamma_H}) + \text{Li}_2(-e^{-\Gamma_H - \gamma_H}) - \text{Li}_2(-e^{-\gamma_H})] - (1 - \eta_C)^2 [g(e^{-\Gamma_C - \gamma_C}) - g(e^{-\gamma_C}) + \text{Li}_2(-e^{-\Gamma_C - \gamma_C}) - \text{Li}_2(-e^{-\gamma_C})] + [(1 - \eta_C)\gamma_H - (1 - \eta_C)^2 \gamma_C][\ln(1 + e^{-\Gamma_C - \gamma_C}) - \ln(1 + e^{-\gamma_C})]\} \quad \text{(B1)}$$

and

$$\mathcal{P} = [\gamma_H - (1 - \eta_C)\gamma_C]\left[(1 - \eta_C)\ln\left(\frac{1 + e^{-\Gamma_C - \gamma_C}}{1 + e^{-\gamma_C}}\right) - \ln\left(\frac{1 + e^{-\Gamma_H - \gamma_H}}{1 + e^{-\gamma_H}}\right)\right] \quad \text{(B2)}$$

respectively, where $g(x) = \ln(x)\ln(1+x)$, $\text{Li}_2(x) = \sum_{i=1}^{\infty} \frac{x^i}{i^2}$, $\gamma_{H/C} = \frac{\varepsilon_l - T_{H/C}}{k_B T_{H/C}}$ and $\Gamma_{H/C} = \frac{\Delta\varepsilon}{k_B T_{H/C}}$. At the given $T_H$, $T_C$ and $\Delta\varepsilon$, the maximum power $\mathcal{P}_m^+$ can be derived with the condition of $\frac{\partial \mathcal{P}}{\partial \gamma_{H/C}} = 0$, and the EMP upper bound $\eta_{\mathcal{P}}^{r+}$ can then be obtained through numerical calculation.

Figure B3 shows more details of ratios ($\mathcal{R}_{\mathcal{P}}$ and $\mathcal{U}_r$) for the same parameters as Figure B2. Combined with Figures 4(c) and (d), the maximum power can be affected significantly by uniformity $\mathcal{U}_r$ and slope $m$ at narrow and wide transmission window, respectively. The reason is due to the variation of $\mathcal{U}_r$ and $m$ with respect to the transmission window. For narrow transmission window, $m$ is large enough, and the difference between ideal boxcar and boxcar-like transmission mainly relies on $\mathcal{U}_r$. However, for wide transmission window, $m$ decreases rapidly, and the transmission tails outside the transmission window destroys the boxcar lineshape dramatically. To identify these statements, the electrons in the energy range of is $\varepsilon_l \leq \varepsilon \leq \varepsilon_u$ are counted for the contribution to the maximum power, i.e. $m \to \infty$, and we found that the variation of the maximum power is very similar to that of the uniformity, seen in Figure B3(c).

From the results of Figures B3(a) and (c), the maximum power can be further optimized with respect to the structure size of the superlattice. Taking advantage of the machine learning technique, we then extract the optimum maximum power with respect to $\Delta\varepsilon$ when $T_C = 10$ K. The optimum bias voltages are not constant with $\Delta\varepsilon$, as shown in Figure 6. Then, the training data is generated with different bias voltages, i.e. $V$ increases from 2.95 mV to 3.15 mV by the step of 0.025 mV, where the corresponding $\Delta\varepsilon$ is still extracted when $b$ and $w$ increase from 3 nm to 8 nm by the step of 0.2 nm. The predicted optimum $b$ and $w$ for optimum maximum power at given bias voltage are then chosen to calculated the maximum power and EMP, see Figure B4(a) and the cyan dots in Figure 7. The corresponding uniformity and slope are shown in Figures B4(b) and (c), which imply again that the maximum power can be affected significantly by the uniformity and the slope at narrow and wide transmission window, respectively.

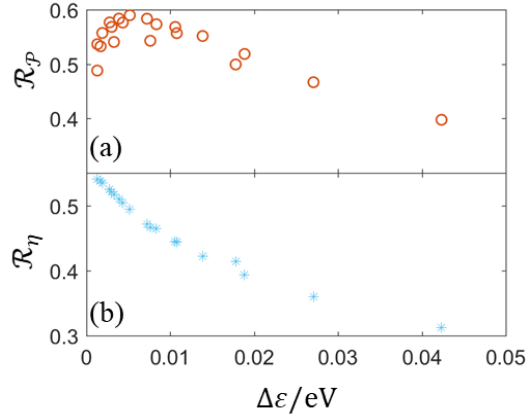

**Figure B1** (a) The maximum power ratio of that from our TEHE to that from ideal boxcar transmission. (b) the corresponding EMP ratio. The parameters are $T_H = 40$ K, $T_C = 10$ K, $V = 1$ mV.

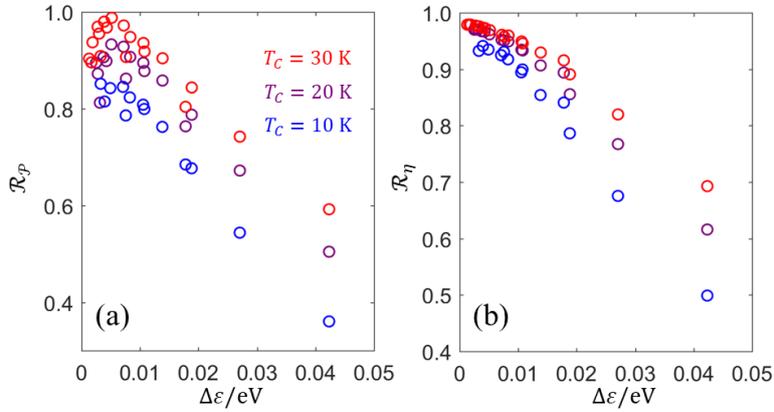

**Figure B2** (a) The maximum power ratios and (b) the EMP ratios with respect to the energy width $\Delta\varepsilon$ of transmission window. The temperature of hot reservoir is fixed as $T_H = 40$ K, and the bias voltage $V = 3$ mV, 2 mV, 1 mV for $T_C = 10$ K, 20 K, 30 K, respectively.

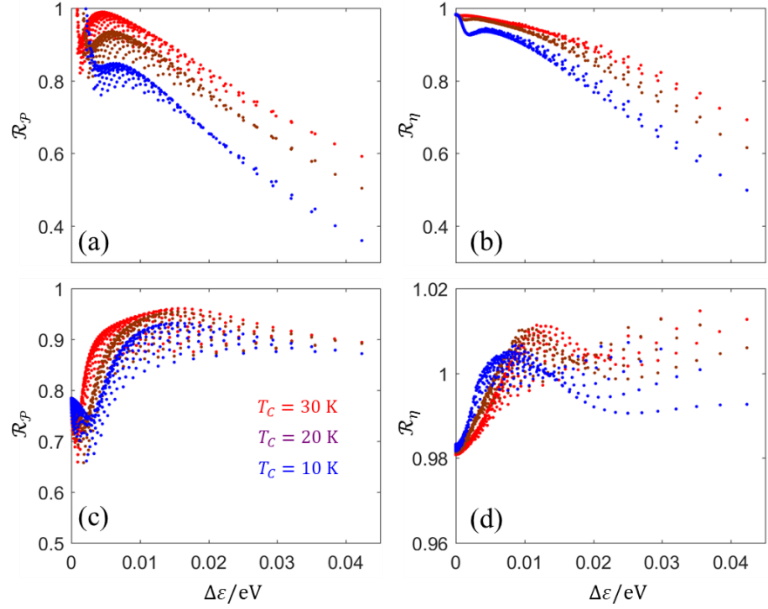

**Figure B3** (a) and (b) are more details of Figure B2, where $\Delta\varepsilon$ is extracted when $b$ and $w$ increase from 3 nm to 8 nm with the step of 0.2 nm. (c) and (d) are the results that counting the electrons in the range of is $\varepsilon_l \leq \varepsilon \leq \varepsilon_u$. Other parameters are the same as Figure B2.

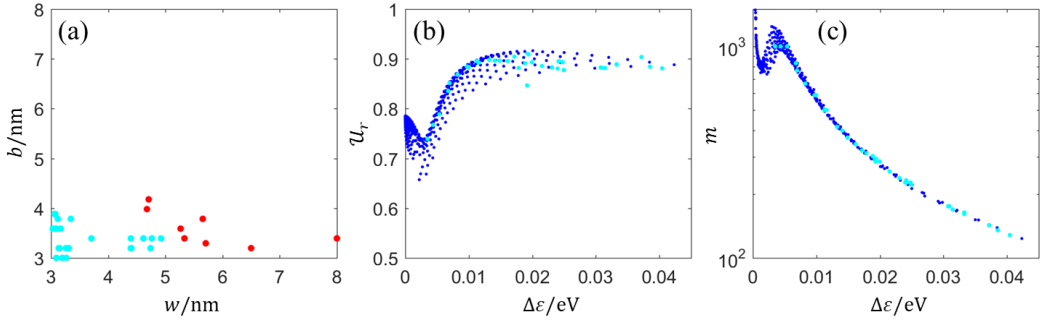

**Figure B4** (a) The predicted optimum widths of barrier and well from the machine leaning results, the red dots denote cases of $\mathcal{R}_\mathcal{P} > 0.8$. (b) The uniformity $\mathcal{U}_r$ and (c) the slope $m$ with respect to the energy width of transmission window for the case of $T_C = 10$ K and $T_H = 40$ K. The blue dots indicate the calculated results when $V = 3$ mV, the cyan dots denote the machine leaning results.